\documentclass[aps,prc,twocolumn,superscriptaddress,showpacs,10pt,dvipsnames]{revtex4-1}

\usepackage[utf8]{inputenc}
\usepackage{amsfonts}
\usepackage{amsmath}
\usepackage{amssymb}
\usepackage[pdftex]{graphicx}
\usepackage{xcolor}
\usepackage{tikz}
\usepackage{cancel}
\usepackage{comment}
\usetikzlibrary{arrows,patterns}

\begin{document}

\newcommand{\tj}[6]{ \begin{pmatrix}
  #1 & #2 & #3 \\
  #4 & #5 & #6
 \end{pmatrix}}


\title{Determining astrophysical three-body radiative capture reaction rates from inclusive Coulomb break-up measurements}




\author{J. Casal}
\email{jcasal@us.es}
\author{M. Rodr\'{\i}guez-Gallardo}
\author{J. M. Arias}
\affiliation{Departamento de F\'{\i}sica At\'omica, Molecular y Nuclear,
  Facultad de F\'{\i}sica, Universidad de Sevilla, Apartado 1065, E-41080
  Sevilla, Spain} 
\author{J. G\'omez-Camacho}
\affiliation{Departamento de F\'{\i}sica At\'omica, Molecular y Nuclear,
  Facultad de F\'{\i}sica, Universidad de Sevilla, Apartado 1065, E-41080
  Sevilla, Spain} 
\affiliation{Centro Nacional de Aceleradores, Universidad de Sevilla, Junta de Andalucía, CSIC, Tomas A Edison 7, E-41092
  Sevilla, Spain} 

\date{\today}

\begin{abstract}
A relationship between the Coulomb inclusive break-up probability and the radiative capture reaction rate for weakly-bound three-body systems is established. This direct link provides a robust procedure to estimate the reaction rate for nuclei of astrophysical interest by measuring inclusive break-up processes at different energies and angles. This might be an advantageous alternative to the determination of reaction rates from the measurement of $B(E1)$ distributions through exclusive Coulomb break-up experiments. In addition, it provides a reference to assess the validity of different theoretical approaches that have been used to calculate reaction rates. 
The procedure is applied to $^{11}$Li ($^{9}$Li+n+n) and $^6$He ($^{4}$He+n+n) three-body systems for which some data exist.

\end{abstract}


\maketitle
Nucleosynthesis occurs in stellar environments, following a complex network of reactions in which heavier nuclei are formed by proton, neutron, or $\alpha$ radiative capture by a lighter nucleus. The knowledge of these radiative capture reaction rates is  crucial for the stellar models aiming to describe the evolution in composition, energy production, and temperature structure of different stellar environments (see, for example, Ref.~\cite{Langanke01} and references therein). The direct experimental measurement of the relevant cross sections is, in principle, possible for two-body reactions~\cite{Iliadis,JPraena13}.
In many interesting cases, however, reaction cross sections cannot be measured directly. This may occur if the initial nucleus is short-lived~\cite{Arnould99} or when the capture process is a three-body reaction~\cite{Sumiyoshi02}. In this case, the inverse reaction to radiative capture, photodissociation, could be measured~\cite{Arnould99}. Reaction rates are then obtained by integrating the photodissociation cross section for the compound nucleus, weighted with the Maxwell-Boltzmann energy distribution and the relevant phase space factors, from the corresponding energy threshold~\cite{Fowler67,RdDiego10}. Direct photodissociation measurements can be done only for stable nuclei, e.g.,~$^{12}$C~\cite{Gai11}, sometimes with important discrepancies among different experiments, e.g.,~$^9$Be~\cite{Sumiyoshi02,Arnold12}. Thus, for many relevant cases this technique is not feasible (e.g.,~$^{17}$Ne~\cite{Marganiec14}). In addition, to obtain the reaction rate from experimental photodissociation measurements usually requires, for three-body systems, a sequential description of the formation process, which is questionable at low temperatures if the particles do not have enough energy to populate intermediate resonances~\cite{Garrido11}.

At first order, the energy distribution of the photodissociation cross section is determined by the $B(E1)$ distribution, of the compound nucleus, into the continuum of its fragments~\cite{Forssen03}. Hence, 
an alternative to obtain this cross section, when the nucleus is short-lived, is to perform exclusive Coulomb break-up  experiments at intermediate energies ($\sim$100 MeV/nucleon) on heavy targets, at very forward angles~\cite{TAumann99,TNakamura06}. From the exclusive break-up cross section, the $B(E1)$ is extracted assuming Coulomb is dominant at those angles. However, these experiments have large uncertainties precisely for the energies close to the threshold, which are the most relevant for the astrophysical reaction rate. The $B(E1)$ distributions could be also calculated, provided that a reliable few-body model is known for the compound nucleus~\cite{RdDiego10,JCasal13,JCasal14}. Nevertheless, different models may provide different results for the same system. Moreover, these calculations do not always agree with the experimental $B(E1)$ distributions. 


Recently, a method to obtain $B(E1)$ distributions close to the break-up threshold has been proposed~\cite{JPFerGar13}, which consists in measuring Coulomb excitation at low energies, i.e.,~around the Coulomb barrier. In this case, one has to measure the {\it inclusive} break-up cross section, which depends mainly on the $B(E1)$ values close to the threshold. The inclusive break-up probability depends on an integral over the $B(E1)$ distribution. That integral is weighted by an exponential factor, which is formally equivalent to the Maxwell-Boltzmann exponential factor in the astrophysical reaction rate. Thus, the $B(E1)$ distribution allows us to establish a correlation between both observables, the break-up probability and the reaction rate. The explicit relation is obtained in this work. In order to show how to implement this method in the case of three-body radiative capture, two examples are worked out: the two-neutron radiative capture by $^9$Li to produce $^{11}$Li and by $^4$He to produce $^6$He. The $^9$Li($2n,\gamma$)$^{11}$Li reaction could appear in the $\alpha$--process in type II supernovae or in the Inhomogeneous Big Bang~\cite{Hoberhummer95}. Although the $^{11}$Li formation might not be very relevant for Astrophysics, this case is chosen to illustrate the method since reasonable experimental data on inclusive break-up cross sections has been measured recently at TRIUMF~\cite{JPFerGar13} at the angles required for the applicability of the present procedure. As a case of more astrophysical interest, the $^4$He($2n,\gamma$)$^{6}$He reaction is presented. This is considered to be relevant for the r-process in neutron star mergers~\cite{Bartlett06}. However, the available data on the break-up angular distribution do not reach the most appropriate angular region. No data exist for other systems of astrophysical interest. If inclusive Coulomb break-up experiments at low energy are performed for these systems in the appropriate angular region, the method here presented will allow us to get, in addition to other important results for Nuclear Physics, the reaction rate of interest in Astrophysics.

The energy-averaged reaction rate for three-body (\mbox{$a+b+c \rightarrow$ $A + \gamma$}) 
radiative capture processes can be calculated from the inverse photodissociation reaction as a function of the temperature  $T$ by the expression~\cite{RdDiego10,Fowler67}
\begin{equation}
 \langle R_3(\varepsilon)\rangle(T) =  \frac{\mathcal{C}'_3 e^{\frac{|\varepsilon_B|}{k_B T}}}{\left(k_B T\right)^{3}}   \int_{|\varepsilon_B|}^\infty d\varepsilon_\gamma~\varepsilon_\gamma^2 \sigma_\gamma(\varepsilon_\gamma) e^{\frac{-\varepsilon_\gamma}{k_B T}},
 \label{eq:aRE}
\end{equation}
where $\varepsilon=\varepsilon_\gamma+\varepsilon_B$ is the initial three-body kinetic energy, $\varepsilon_\gamma$ is the energy of the photon emitted, and $\varepsilon_B$ is the binding energy of the compound nucleus $A$. Here, $\mathcal{C}_3'$ is a constant depending on the masses and spins of the initial particles. The photodissociation cross section $\sigma_\gamma(\varepsilon_\gamma)$ of the nucleus $A$ can be expanded into electric and magnetic multipoles, $\lambda$, as~\cite{Forssen03,JCasal14}
\begin{equation}
\sigma_\gamma^{(\lambda)}(\varepsilon_\gamma)=\frac{(2\pi)^3 (\lambda+1)}
{\lambda[(2\lambda+1)!!]^2}\left(\frac{\varepsilon_\gamma}{\hbar
c}\right)^{2\lambda-1}\frac{dB({\cal O}\lambda)}{d\varepsilon},
\label{eq:xsection}
\end{equation}
where $B({\cal O}\lambda)$ is the order $\lambda$ of electric or magnetic transition probability (${\cal O} = E, M$).
Hence, in leading order, the astrophysical reaction rate is due to the dipole electric contribution ($E1$) and can be written as 
\begin{equation}
  \langle R_3(\varepsilon)\rangle(T) \simeq \frac{\mathcal{C}_3 e^{\frac{|\varepsilon_B|}{k_B T}}}{\left(k_B T\right)^{3}}   \int_{|\varepsilon_B|}^\infty d\varepsilon_\gamma~\varepsilon_\gamma^3 \frac{dB(E1)}{d\varepsilon} e^{\frac{-\varepsilon_\gamma}{k_B T}}.
 \label{eq:aRE_be1}
\end{equation} 
The explicit form of the constant $\mathcal{C}_3$ in Eq.~(\ref{eq:aRE_be1}) is
\begin{equation}
\mathcal{C}_3 = \nu!\frac{\hbar^2}{c^3}\frac{2^7\pi^4}{3^2(a_x a_y)^{3/2}}\frac{g_A}{g_a g_b g_c},
\end{equation}
where $g_i$ are the spin degeneracies of the particles, $\nu$ is the number of identical particles in the three-body system, and $a_x$ and $a_y$ are the reduced masses of the subsystems related to the Jacobi coordinates $\{\boldsymbol{x},\boldsymbol{y}\}$.

In this work we propose an alternative way to determine the astrophysical reaction rate taking advantage of the sensitivity to Coulomb excitation in scattering processes at low energy. 
At energies around the Coulomb barrier and at very forward angles, the break-up probability of weakly-bound nuclei also depends, in leading order, on the $B(E1)$ distribution. In the Equivalent Photon Method (EPM), the reduced break-up probability in the center-of-mass frame is written as~\cite{JPFerGar13}
\begin{equation}
  P_{\rm{r} }(t) = t^2 ~ \int_{|\varepsilon_B|}^\infty d\varepsilon_\gamma~\varepsilon_\gamma~ \frac{dB(E1)}{d\varepsilon}  e^{-t\varepsilon_\gamma} ,
  \label{eq:Pbu-red}
\end{equation}
where, $\varepsilon_\gamma$ and $\varepsilon_B$ have the same meaning as above and  $t$ is the collision time, which is related to the scattering angle in the center-of-mass frame through
\begin{equation}
 t=\frac{a_0}{\hbar v}\left(\pi+\frac{2}{\sin(\theta/2)}\right).
 \label{eq:tt}
\end{equation}
Here, $a_0$ is half the classical closest approach distance and $v$ is the relative projectile-target velocity. With this definition, the collision time, $t$, is independent of the collision parameters. This makes $t$ a scaling variable in such a way that experiments at different energies can be merged together and analyzed with a single quantity, $P_{\rm r}(t)$~\cite{JPFerGar13}.
Equations~(\ref{eq:aRE_be1}) and (\ref{eq:Pbu-red}) are formally equivalent except for a factor $\varepsilon_\gamma^2$. This means that both observables, reaction rate and break-up probability, are strongly correlated in the range for which the EPM holds. The EPM is valid when the semiclassical approach is valid ($\eta \gg 1$), nuclear forces are negligible ($\theta < \theta_g$), and Coulomb coupling can be approximated by first order, so that higher order Coulomb excitations are negligible. This opens the possibility of getting reliable information on the astrophysical reaction rate from experimental measurements of the break-up probability.
The maximum correlation between both observables is established when the exponentials in Eqs.~(\ref{eq:aRE_be1}) and (\ref{eq:Pbu-red}) are equal, i.e.
\begin{equation}
 t = \frac{1}{k_B T},
 \label{eq:ttT}
\end{equation}
which, together with Eq.~(\ref{eq:tt}), establishes a direct correspondence between the scattering angle, $\theta$, and the temperature, $T$. This relation is not unique since it depends on the bombarding energy through the parameter $a_0$ and the velocity $v$.
This fact opens the possibility of exploring different temperature ranges of relevance in Astrophysics by measuring break-up probabilities at different energies. For smaller energies, one gets information for the same angle on larger collision times, which corresponds to exploring lower temperatures.

Note that Eqs.~(\ref{eq:aRE_be1}) and~(\ref{eq:Pbu-red}) can be related through
\begin{equation}
\langle R_3(\varepsilon)\rangle(T) = C_3 t^{3}e^{|\varepsilon_B|t} \frac{d^2}{dt^2}\left(\frac{1}{t^2}P_r(t)\right). 
\label{eq:RP}
\end{equation}
Equation (\ref{eq:RP}) is the main result of this work. It relates directly the reaction rate in a stellar environment at a given temperature $T$ with the inclusive break-up probability $P_r(t)$ obtained in a Coulomb scattering experiment, for certain collision times corresponding to given scattering angles and energies. 

\begin{figure}
\includegraphics[width=0.45\textwidth]{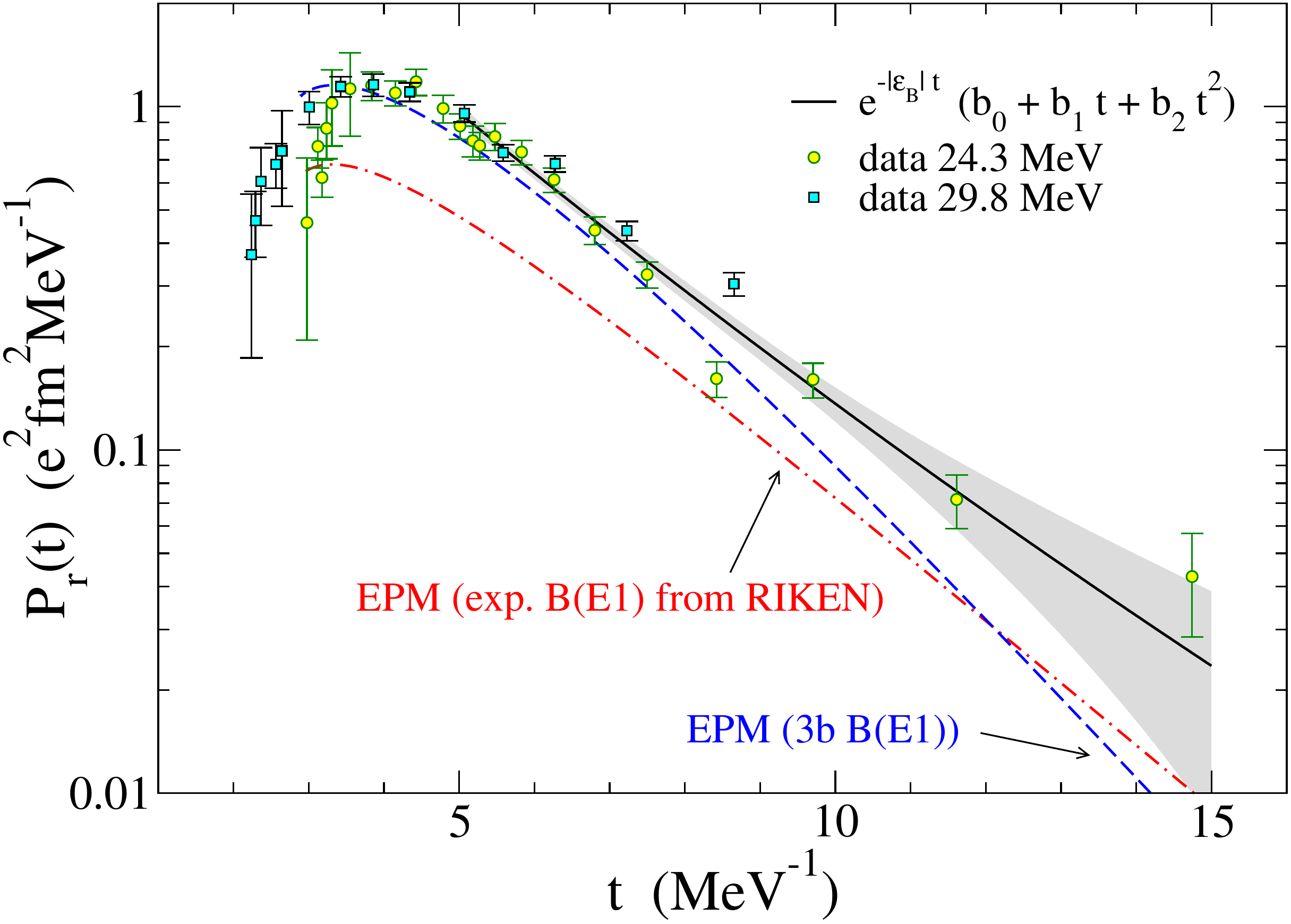}
\caption{(Color Online) $^{11}$Li + $^{208}$Pb reduced break-up probability as a function of the collision time, $t$. See text for details.} 
\label{fig:Pred} 
\end{figure}

To evaluate from a practical purpose the second derivative in Eq.~(\ref{eq:RP}), it is convenient to fit a suitable function to the experimental data.
It is found (see Fig.~\ref{fig:Pred}) that the main $t$--dependence of $P_r(t)$ is through the exponential factor $e^{-|\varepsilon_B|t}$, as can be deduced from Eq.~(\ref{eq:Pbu-red}). Thus, without any loss of generality, we can express $P_r(t)$ as an expansion,
\begin{equation}
P_r(t)\simeq e^{-|\varepsilon_B|t} \left(b_0+b_1 t+b_2 t^2+\dots \right).
\label{eq:Pbu-fit}
\end{equation}
The parameters $b_0,b_1,b_2,\dots$ are fitted to the experimental values of $P_r(t)e^{|\varepsilon_B|t}$, over the range which is Coulomb dominated. From these values, we can obtain through Eq.~(\ref{eq:RP}) an expansion of the reaction rate,
\begin{eqnarray}
\nonumber\langle R_{3}(\varepsilon) \rangle(T)=& C_3 & \left[b_0\left(|\varepsilon_B|^2k_BT+4|\varepsilon_B|+6/(k_BT)\right)\right. \\\nonumber & + & b_1\left(|\varepsilon_B|^2(k_BT)^2 + 2|\varepsilon_B|k_BT + 2\right) \\ & + & b_2\left(|\varepsilon_B|^2(k_BT)^3\right) + \dots\left.\right].
\label{eq:ratett}
\end{eqnarray}
The parameters $b_0,b_1,b_2,\dots$ fitted in Eq.~(\ref{eq:Pbu-fit}) will have some uncertainties, given by a covariance matrix. These uncertainties can be implemented in Eq.~(\ref{eq:ratett}) to get an error estimate in the reaction rate.

To illustrate this method, we use recent data of $^{11}$Li break-up on $^{208}$Pb at low energies~\cite{JPFerGar13}. In Fig.~\ref{fig:Pred} we present the $^{11}$Li + $^{208}$Pb reduced break-up probability. 
We see that, in the region from $t=5$ to $15$  MeV$^{-1}$, data are reasonably smooth and follow an exponential decay. We fit the product of the break-up probability $P_r(t)$ times $e^{|\varepsilon_B|t}$ by a second-degree polynomial, Eq.~(\ref{eq:Pbu-fit}), obtaining the values $b_0=7.8$ e$^2$fm$^2$MeV$^{-1}$, $b_1=-0.4$ e$^2$fm$^2$, $b_2=0.02$ e$^2$fm$^2$MeV, with the corresponding covariance matrix. 
In this fit, $|\varepsilon_B|$ has been taken as 0.37 MeV, the experimental two-neutron separation energy of $^{11}$Li~\cite{Smith08}. In Fig.~\ref{fig:Pred} the solid black line is the result of the quadratic fit and the shadow region around is the 1--$\sigma$ uncertainty region. For comparison, in Fig.~\ref{fig:Pred} we also include the results obtained by integrating directly through Eq~(\ref{eq:Pbu-red}): (i) the experimental $B(E1)$ distribution~\cite{TNakamura06} (dot-dashed red line) and (ii) a theoretical three-body $B(E1)$ distribution~\cite{JPFerGar13} (dashed blue line). It is shown that the quadratic fit reproduces fairly well the experimental data on the break-up probability.

\begin{figure}
\includegraphics[width=0.45\textwidth]{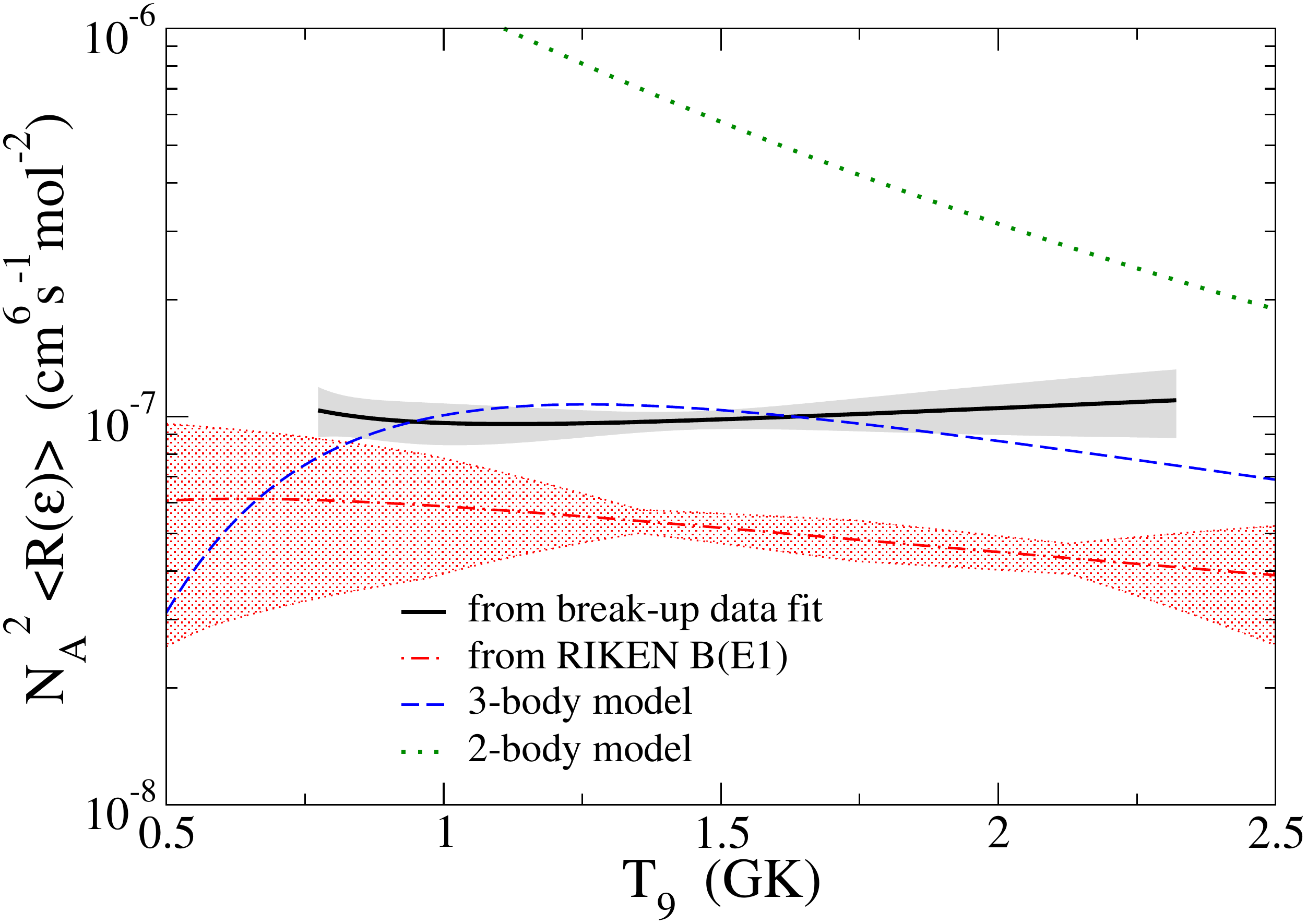}
\caption{(Color online) Reaction rate for $^{11}$Li formation as a function of temperature (in GK). See text for details.} 
\label{fig:rate}
\end{figure} 

From the knowledge of parameters $b_0$, $b_1$ and $b_2$, we can predict the reaction rate (in the range of temperatures from 0.7 to 2.3 GK, corresponding to collision times from $t=15$ to $t=5$ MeV$^{-1}$) for the two-neutron capture by $^9$Li using Eq.~(\ref{eq:ratett}), and propagate the 1--$\sigma$ band of uncertainty from the fit. 

In Fig.~\ref{fig:rate} we show the reaction rate as a function of the temperature. The result from the quadratic fit and its corresponding 1--$\sigma$ uncertainty band are shown by the solid black line and the shaded region, respectively. In the same figure, the results obtained from different $B(E1)$ energy distributions are also shown: (i) the experimental RIKEN data~\cite{TNakamura06} (dot-dashed red line), (ii) a theoretical three-body model of $^{11}$Li which presents a dipolar resonance at 0.69 MeV~\cite{JPFerGar13} (dashed blue line), and (iii) a theoretical two-body model of $^{11}$Li with a dipolar resonance at the same position~\cite{LayTh} (dotted green line). The result obtained from the RIKEN experimental data includes an estimate of the uncertainty in the reaction rate, which is due to the statistical uncertainties of the $B(E1)$ points and also to the uncertainty in the break-up energy. The latter is given \cite{TNakamura06} as $\Delta E = 0.17 \sqrt{\varepsilon}$, with $\varepsilon$ in MeV. This uncertainty is especially important, in relative terms, for the energies close to the threshold, which are the most relevant for the reaction rate. As shown in Fig.~\ref{fig:rate}, the reaction rate extracted from the present work is significantly larger than the reaction rate extracted from the RIKEN $B(E1)$ values, although these have considerable uncertainties. The difference is also seen in the break-up probability (Fig.~\ref{fig:Pred}). 
This discrepancy should be further investigated since it can be due to the validity of EPM, used in both approaches, or the treatment of the exclusive experimental data to obtain the $B(E1)$ distribution. 
On the other hand, it is seen in Fig.~\ref{fig:rate} that the reaction rate obtained from the three-body $B(E1)$ distribution in Ref.~\cite{JPFerGar13} is in good agreement with the present estimate from inclusive break-up data, for the region of temperatures corresponding to the collision times explored in this experiment. In contrast, the two-body calculation~\cite{LayTh}, including a resonance at the same energy as in the three-body model, is clearly out of the values and trend obtained by the method here presented. 

The presented method can be applied to estimate the reaction rate of the two-neutron capture on $^4$He to produce $^6$He, which has been proposed to affect the r-process nucleosynthesis in neutron star mergers~\cite{Bartlett06}. Several estimations of this reaction rate have been carried out~\cite{Bartlett06,RdDiego10,JCasal13}, showing noticeable differences between them. 

\begin{figure}
\includegraphics[width=0.45\textwidth]{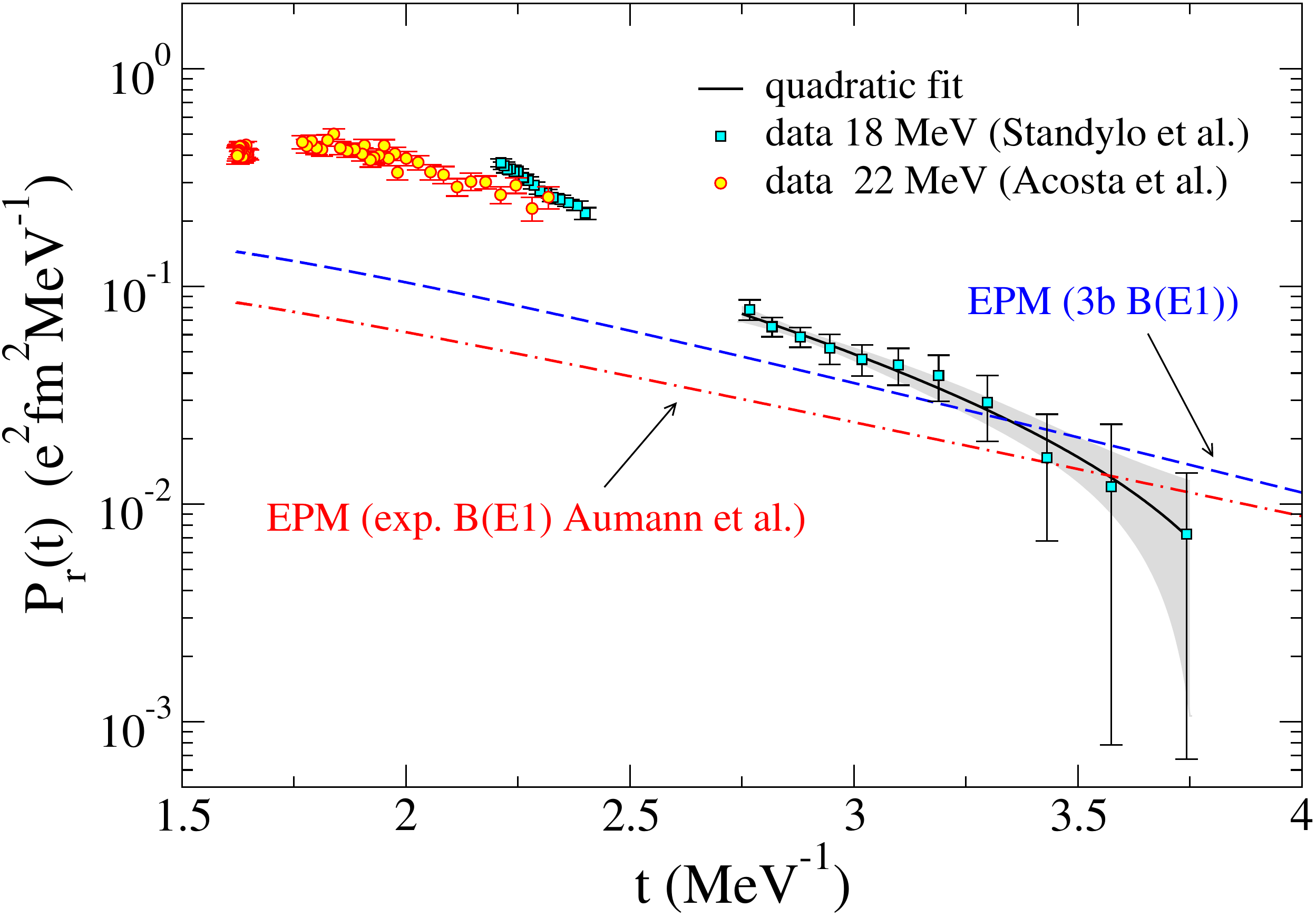}
\caption{(Color Online) $^{6}$He + Pb reduced break-up probability as a function of the collision time, $t$. See text for details.}
\label{fig:Pred_6he} 
\end{figure}

Two sets of experimental data on $^6$He inclusive break-up are available in the literature: the scattering on $^{208}$Pb at 22 MeV~\cite{Acosta11}, and the reaction on $^{206}$Pb at 18 MeV~\cite{Standylo13}. However, none of them explore the small angle region in which the EPM approach is valid. Nevertheless, the present method can be applied at the expense of introducing a systematic error, which will be related to nuclear and higher order effects. In Fig.~\ref{fig:Pred_6he} we present the $^6$He + Pb reduced break-up probability corresponding to the two data sets. To reduce the systematic error, we can perform the quadratic fit considering only the data at relatively larger collision times. This is given by parameters $b_0=2.01$ e$^2$fm$^2$MeV$^{-1}$, $b_1=-0.03$ e$^2$fm$^2$ and $b_2=-0.13$ e$^2$fm$^2$MeV. In this fit, $|\varepsilon_B|$ has been taken as 0.975 MeV, the experimental two-neutron separation energy of $^{6}$He~\cite{Tilley02}. In Fig.~\ref{fig:Pred_6he} the solid black line is the result of the quadratic fit and the shadow region around is the 1--$\sigma$ uncertainty region. We also include in Fig.~\ref{fig:Pred_6he} the result obtained by integrating the experimental $B(E1)$ from~\cite{TAumann99} (dot-dashed red line) and a theoretical three-body $B(E1)$~\cite{JCasal13} (dashed blue line).

\begin{figure}
\includegraphics[width=0.45\textwidth]{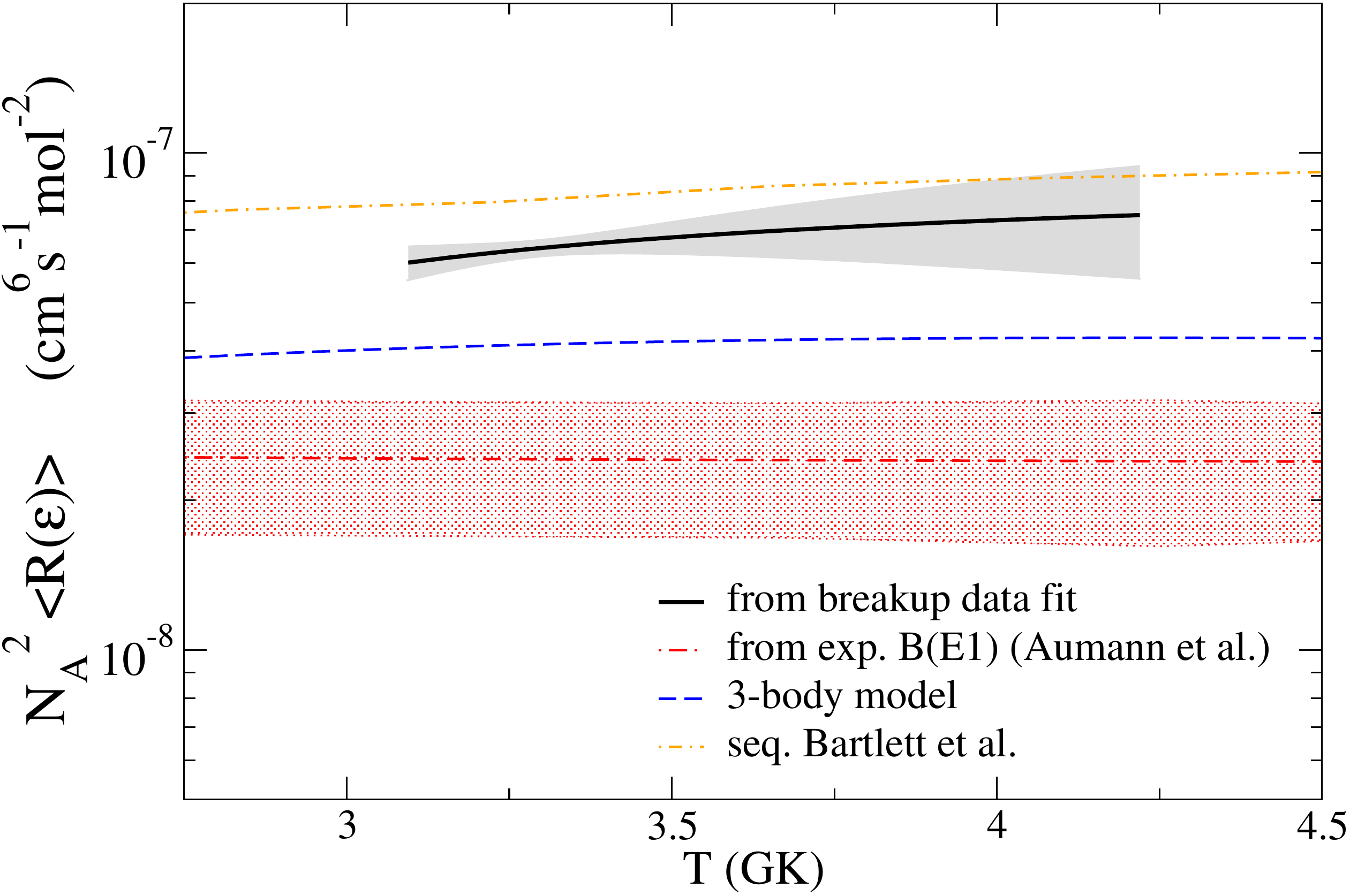}
\caption{(Color online) Reaction rate for $^{6}$He formation as a function of temperature (in GK). See text for details.} 
\label{fig:rate_6he}
\end{figure}

The reaction rate obtained from the knowledge of parameters $b_0,b_1$ and $b_2$ is shown in Fig.~\ref{fig:rate_6he} (solid black line). The temperature range covered by the data goes from 3.1 to 4.2 GK. The present estimate lays between the sequential calculation from~\cite{Bartlett06} (dot-dashed green line) and the full three-body calculation in Ref.~\cite{JCasal13} (dashed blue line). As in the case of $^{11}$Li, the estimation from the experimental $B(E1)$~\cite{TAumann99} provides a smaller reaction rate in the whole temperature range. It should be remarked that the experimental data used for the fit cover an angular region between 40$^\circ$ and 70$^\circ$ (which, for this reaction, corresponds to collision times between 3.74 and 2.77 MeV$^{-1}$), where corrections to the EPM are expected to change the break-up cross sections. Experimental data on $^6$He break-up at larger collision times would allow us to minimize the systematic error and to explore lower astrophysical temperatures. As a reference, for the reaction ${^6}$He+$^{206}$Pb at $E_{lab}$=18 MeV, if the reaction rate is needed at around $T=1$ GK break-up measurements starting on $\theta_{min}=9.5^{\circ}$ (corresponding to a collision time of 11.6 MeV$^{-1}$) have to be performed. 


In summary, we have established a relation between the radiative capture reaction rate and the inclusive Coulomb break-up probability in the region in which first-order dipole Coulomb interaction is dominant. It should be noted that the present formalism is worked out here for three-body systems but its extension to two-body systems is straightforward and will be presented elsewhere. The temperature of the stellar environment is directly related to the collision time of the reaction, which depends on the scattering angle and the incident energy. This implies that detailed measurements on inclusive break-up probabilities of these systems will provide a direct estimation of the corresponding reaction rates of astrophysical interest in a given range of temperatures. This result establishes a new experimental tool to determine astrophysical reaction rates for short-lived nuclei and provides an additional motivation to carry out Coulomb scattering experiments involving exotic nuclei in radioactive ion beam facilities. There are very few experimental data on inclusive break-up scattering at energies around the Coulomb barrier for nuclei of astrophysical interest. Some three-body capture processes of astrophysical interest whose reaction rates could be estimated with the present method are $\alpha(\alpha n,\gamma){^{9}\text{Be}}$ or $^{15}\text{O}(2p,\gamma){^{17}\text{Ne}}$. The present formalism implies that inclusive Coulomb break-up measurements, in addition to the nuclear structure information, will provide an extra impact in the Nuclear Astrophysics field. 

\begin{acknowledgments}
Authors are grateful to J. P. Fernández-García and J. A. Lay for useful discussions and suggestions. 
This work has been partially supported by the Spanish Ministerio de Econom\'{\i}a y Competitividad and the European Regional Development Fund (FEDER) under Projects FIS2011-28738-c02-01, FIS2013-41994-P, FPA2013-47327-C2-1-R, FIS2014-53448-c2-1-P and FIS2014-51941-P, and by Junta de Andaluc\'{\i}a under group number FQM-160 and Project P11-FQM-7632. 
J. Casal acknowledges support from the Ministerio de Educaci\'on, Cultura y Deporte, FPU Research Grant AP2010-3124. M. Rodríguez-Gallardo acknowledges postdoctoral support from the V Plan Propio de Investigación of the Universidad de Sevilla, contract number USE-11206-M.
\end{acknowledgments}


\bibliography{./bibfile}

\end{document}